\documentstyle[floats,aps,prl,twocolumn,epsf]{revtex}

\begin{document}
\draft
\wideabs{
\title{Coulomb Effects in Spectral Density and Transverse
Conductivity of Layered Metals}
\author{S.N.Artemenko and S.V.Remizov}
\address{
Institute for Radioengineering and Electronics of Russian Academy
of Sciences, Mokhovaya 11/7, 101999 Moscow, Russia }

\date{\today}
\maketitle
\begin{abstract}
Renormalization of the Coulomb interaction in layered metals
results in a strongly anisotropic plasma mode with low frequencies
for small components of wave vector in the in-plane direction.
Interaction of electrons with this mode was found to lead to an
incoherent contribution to the electron spectral density which
spreads up to large energies. In the superconducting state this
reproduces the peak-dip-hump feature similar to that observed in
layered high-T$_c$ superconductors. The incoherent part of the
spectral density and plasmon-assisted electron transitions provide
mechanisms for nearly linear conductivity in the stack direction
at large voltages or frequencies.
\end{abstract}
}

Layered crystals can be considered as very anisotropic crystals
with small interlayer transfer integral determining the electronic
bandwidth in the transverse direction. When such a crystal is pure
enough so that its periodicity is not strongly disturbed the
interlayer electron transitions are expected to be coherent, {\it
i. e.} the in-plane component of the momentum is conserved, in
contrast to rough artificial tunnel junctions where it is not
conserved. As it is well-known from the textbooks, electrons
oscillate in an electric field applied to an ideal crystal, and do
not produce the dissipative current. A finite resistivity (and the
real part of conductivity) in the stack direction appears due to
electron scattering. At frequencies exceeding the in-layer
scattering rate, $1/\tau$, the scattering becomes ineffective and
conductivity is expected to decrease as frequency increases.
Similarly, the time needed for an electron to oscillate in the
electric field becomes smaller than $\tau$ if the voltage drop per
a single layer becomes larger than $\hbar/e\tau$, and the real
part of the conductivity must decrease with voltage increasing.
However, one of the most studied layered metals, high-T$_c$
superconductors, exhibit finite conductivity with nearly linear IV
curves even at voltages much larger than the superconducting gap
\cite{LB,Kleiner,Yurgens}, which in turn exceeds $\hbar/\tau$. The
real, dissipative part of the conductivity does not disappear at
large frequencies as well \cite{freq}. Similar behavior was
observed in layered metal 2H-TaSe$_2$ \cite{Ta}.

Furthermore, the spectral density for electrons in high-T$_c$
superconductors measured by means of angle-resolved photoemission
spectroscopy contains the peak-dip-hump structure with an
incoherent part spreading up to energies far from the Fermi
surface \cite{ARPES}. The similar behaviour was observed also by
means of tunneling spectroscopy \cite{Huang}. The presence of the
incoherent part may account for the finite dissipative current at
large frequencies \cite{IM}. As it was shown by Norman et al.
\cite{Norman} the features in the spectral density may be modeled
by coupling of electrons to some non-dispersive bosonic mode,
presumably, of electronic origin. An alternative explanation was
presented by Chubukov and Morr \cite{Chub} who argued that the
unusual superconducting properties of cuprates could be explained
by a strong interaction between electrons and overdamped spin
fluctuations peaked at some momentum.

We suggest another mechanism for the discussed properties that is
related to general properties of layered metals and is not
confined to specific properties of high-T$_c$ superconductors. We
calculate the effect of Coulomb interaction on the electron
spectral density and conductivity at large voltages and
frequencies, and find that discussed features of layered metals
can be induced by a strongly anisotropic plasma mode interacting
with electrons. Such a mode must be inherent to layered materials
because frequency of plasma oscillations with wave vector
perpendicular to conducting layers is proportional to the square
of the interlayer transfer integral and, hence, to the transverse
conductivity, the latter being small in highly anisotropic layered
materials. The Josephson plasma mode in high-T$_c$ superconductors
\cite{plth,plex} can be considered as a manifestation of such a
mode in the superconducting state. It is evident that similar mode
must exist also in the normal state at frequencies larger than the
scattering rate and at wave lengths shorter than mean-free-path,
{\it i. e.} in the limit where a metal behaves as an ideal
conductor, its response to electric field being in many respects
similar to that of a superconductor.

We consider a layered metal with the lattice period $s$ in the
direction perpendicular to the metallic layers coupled by small
transfer integral $t_{\perp}$. The Hamiltonian of the system is
\begin{equation}
\begin{array}{c}
\displaystyle
{\cal H} =
\sum_{{\bf p},n,\sigma}
\Bigl[
  \frac{p^2}{2m} a^+_{{\bf p}n\sigma}a_{{\bf p}n\sigma}+
  t_{\perp}(
    a^+_{{\bf p},n+1,\sigma}a_{{\bf p}n\sigma} +
\\
\displaystyle
    a^+_{{\bf p},n-1,\sigma}a_{{\bf p}n\sigma}
  ) +
\Bigr]
+ {\cal H}_{C}+{\cal H}_{BCS}
\end{array}
,
\end{equation}
where $a^+_{{\bf p},n,\sigma}$ is creation operator of electron
with momentum ${\bf p}$ in the plane in the conducting layer $n$
and with spin $\sigma$, $m$ is the in-plane effective mass,
${\cal H}_{C}$ describes Coulomb interaction and ${\cal H}_{BCS}$
is the part of the Hamiltonian of BCS type which leads to singlet
s- or d-wave pairing. We consider a discrete model in which the
potential of Coulomb interaction of two electrons in layers $n$
and $n'$ separated by the in-plane distance $r_\|$ also has a
discrete form $V_{C}=e^2/\sqrt{r_\|^2 +(n-n')^2s^2}.$ Its Fourier
transform reads
\begin{equation}\label{Coul}
V_{C}=4\pi e^2/(q_\|^2+\hat q_\perp ^2),
\end{equation}
where $q_\|$ is the in-plane wave vector, $\hat q_\perp =
(2/s)\sin{(q_\perp s/2)}$ and $|q_\perp|<\pi/s$ is the wave vector
obtained from the discrete Fourier transformation with respect to
layer numbers.

We neglect scattering of the electrons by impurities and phonons
since typical energies involved in our study are large in
comparison to $\hbar/\tau$. We also ignore the regular Coulomb
effects like renormalization of the effective mass and broadening
of the quasiparticle peak in the electronic spectral function due
to finite life-time of the quasiparticles. Instead we concentrate
on the effects induced by plasma mode and its interaction with
electrons.

We calculate first the polarization operator within the RPA
approximation adopted for the anisotropic case and find the
renormalized Coulomb interaction, ${\cal V}_{C}$ at zero
temperature. In the normal state and in the dynamic limit $\omega
\gg q_\| v_F$ it reads
\begin{equation}\label{V}
{\cal V}_{C}(\omega,{\bf q})=
\frac
  {4\pi e^2}
  {(q_\|^2+\hat q_\perp^2)}
\frac
  {\omega^2}
  {\omega^2 -\omega^2_{\bf q}}
,
\quad
\omega^2_{\bf q}
=
\frac
  {\omega^2_{p} \hat q_\perp ^2 + \Omega^2_{p} q_\|^2}
  {q_\|^2+\hat q_\perp ^2}
,
\end{equation}
where $\Omega^2_{p}=2e^2v_Fp_F/s= 4\pi e^2 n/m$ is the in-plane
plasma frequency and $\omega_{p} = (4t_\perp s/\hbar
v_F)\Omega_{p} \ll \Omega_{p}$ is the plasma frequency for wave
vectors normal to the layers. Poles of the renormalized potential
describe the spectrum of the plasma mode. The frequency of the
mode starts from $\omega_{p}$ and in a wide range of frequencies
between $\omega_{p}$ and $\Omega_{p}$ almost linearly increases
with $q_\|$, $\omega_{\bf q} \approx q_\| \Omega_{p}/\hat
q_\perp$. In the calculations below the leading contribution to
integrals comes from large values of $q_\perp \sim s^{-1}$. Thus
the mode can be characterized by typical velocity values $\tilde c
\sim (\kappa s)v_F$, where $\kappa^2 = \Omega_p^2/2v_F^2$,
$1/\kappa$ is the Thomas-Fermi screening length. To simplify our
calculations we assume that this length is smaller than the
spacing $s$, $\kappa s \gg 1$. This relation holds, {\it e. g.},
for high-T$_c$ materials in which $\kappa \sim 10$ nm$^{-1}$ and
$s \sim 1.5$ nm. Note that $\tilde c \gg v_F$, and, hence, we can
use dynamic limit in (\ref{V}).

There is some analogy between interaction of electrons with
photons in quantum electrodynamics and interaction of
electrons with plasmons in our problem. But while in the former
case interaction was described by the small dimensionless
parameter $e^2/(\hbar c) = 1/137$ in the latter case a
corresponding parameter $e^2/(\hbar \tilde c) \sim (e^2/s)/(\hbar
\Omega_p)$ can be either small or not depending on the plasma
frequency and, hence, on the density of free charge carriers. We
consider first the case of a relatively good metal with large
carrier density when this parameter is small. Then we discuss what
happens at smaller carrier concentration.

The renormalized Coulomb potential for the superconducting case in
the quasiclassic limit, $\omega, qv_F \ll \Delta$, coincides with
the potential in the normal state in the static limit, ${\cal
V}_{C}= 4\pi e^2/(q_\|^2+\hat q_\perp ^2+\kappa^2)$. Thus in the
low-frequency limit in the superconducting state there is no
singularities due to the plasma mode in the Coulomb potential.
They recover only at $\omega, qv_F \gg \Delta$ when the
superconductivity is not important. However, the Josephson plasma
mode at low frequencies with the spectrum similar to (\ref{V})
does exist and is manifested by poles of the Green's functions
describing fluctuations of the superconducting momentum (which is
related to the vector potential). These fluctuations also
contribute to the effects we study here, but we do not study them
here because their contribution was found to be by $(\kappa s)^2$
times smaller, than that from the Coulomb effects.

Now we calculate the self-energy of electrons due to renormalized
Coulomb interaction which acquires properties of Green's function
of bosons. The leading contribution to the mass operator is
related to the poles of ${\cal V}_{C}$ describing the plasmons.
This contribution comes from region of large energies. The
resulting mass operator does not decrease up to very large
energies of the order of $\Omega_p$. Using the calculated  mass
operator we find the electronic Green's function in the normal
state at energies much smaller than $\Omega_p$
\begin{equation}\label{G}
\begin{array}{c}
\displaystyle
G=
\frac{1}
  {(\varepsilon - \xi_p)[1+g_0+igF(\varepsilon -\xi_p)/\Omega_p]},
\\
\displaystyle
\quad g=\frac{\pi e^2}{\hbar s \Omega_p},
\quad g_0 = g \int_0^\infty \frac{d{\bf q}\;s \Omega_p}{2\pi^3
\omega_{\bf q} (q_\|^2+ \hat q_\perp^2)},
\end{array}
\end{equation}
where $\xi_p=p^2/2m -\varepsilon_F$ and $F$- is a slowly varying
function equal to 1 at $\varepsilon \ll \xi \kappa s$ and to 1/2
at $\varepsilon \gg \xi \kappa s$. We keep the second small term
in the denominator because it determines the imaginary part of the
Green's function at $\varepsilon \neq \xi_p$. Formally the
integral for $g_0$ logarythmically diverges at large ${\bf q_\|}$.
But it is finite if we take into account that integration over
$q_\|$ is limited by the Brillouin zone or consider the exact
behavior of $\omega_{\bf q}$ at $q_\| \sim \kappa$. Then one gets
$g_0 \sim g \ln{sq_F}$ or $g_0 \sim g \ln{\kappa s}$,
respectively, more detailed expression for $g_0$ being dependent
on the energy structure of the metal at energies far from the
Fermi surface.

Expression (\ref{G}) implies that the spectral function for
electrons, $A=\Im G/\pi$, consists of the suppressed quasiparticle
peak decreasing with decrease of the carrier density and of a
broad feature extending up to the in-layer plasma frequency. For
$\varepsilon \ll \Omega_p$ it has a form
\begin{equation}\label{A}
A=\frac{\delta \,(\varepsilon - \xi_p)}{1+g_0}+ \frac{gF}{
\pi
\Omega_p},
\end{equation}
where we neglected broadening of the quasiparticle peak.

In the superconducting state the expression (\ref{A}) is valid at
$\varepsilon \gg \Delta (\kappa s)$, while at $\varepsilon \ll
\Delta$ the second term in the spectral function is small because
it is determined by poles of the Coulomb potential, and the latter
are not present at energies smaller than $\Delta$. Thus the
spectral density has a dip at energies $\varepsilon < \Delta
(\kappa s)$. Thus there is a peak-dip-hump structure similar to
that observed in high-T$_c$ superconductors in $(0,\pi)$
direction, which corresponds to the maximum of the superconducting
gap. However our analysis within the simple model with the
isotropic electronic spectrum in the conducting plane cannot give
a quantitative description of the spectral density including the
anisotropy related to the d-wave symmetry of the superconducting
order parameter.

Note that though formally equations (\ref{G}-\ref{A}) are valid at
$T=0$ calculation at finite temperature shows that they can be
used at temperatures smaller than plasmon frequencies which give
leading contribution to the integral in (\ref{G}), {\it i. e.} at
$kT \ll \hbar v_F/s,\,\varepsilon_F, \,\Omega_p$. So, practically,
the expression for Green's function is valid at any temperature.

Let us discuss now what happens at larger values of the parameter
$g$ describing the interaction strength. In this case we must
solve the Dyson equations using renormalized Green's function. At
this step one must take into account the vertex correction
$\Gamma$. Vertex diagrams similar to mass operator contain large
contributions from large energies due to plasma-mode poles in the
renormalized Coulomb potential. However our analysis shows that
the vertex corrections do not lead to qualitative changes and with
the accuracy of constant factors equation (\ref{G}) survives if
expressions for $g$ and $g_0$ are assumed to contain the bare
plasma frequency. But the actual plasma frequency is renormalized,
and, hence, depends on the carrier density in different way,
$\Omega_p = \Omega_{p0} /(1 + g_0), \, \omega_{\bf q} \to
\omega_{{\bf q}0}/(1 + g_0)$, where subscript 0 is related to bare
quantities.

To calculate current in the stack direction at large voltages we
use non-equilibrium Green's functions in Keldysh diagram
technique. Current density  between the layers $n$ and $n+1$ is
expressed via the Green's functions which are off-diagonal with
respect to the layer indices:
\begin{equation}
j_{n,n+1}=\int \frac{2e t_\perp}{\hbar} ( G^{12}_{n\,n+1} -
G^{21}_{n+1\,n})\frac{d\varepsilon d{\bf p_\|}}{2\pi^3}, \label{j}
\end{equation}
where upper indices are related to the time contour, and lower
indices describe the layer number.

\begin{figure}[t]
\epsfxsize=8cm \centerline{\epsffile{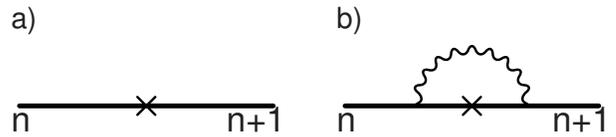}}
\caption{ Diagrams for non-diagonal with respect to layer number Green's
functions, which give main contribution into a current in the stack direction.
} \label{fig1}
\end{figure}
Diagrams describing two basic mechanisms of the origin of linear
conductivity at high voltages are presented in Fig.1. Solid and
wavy lines denote renormalized Green's function and Coulomb
potential, respectively, and cross denotes $t_{\perp}$. These
diagrams give strict results in the second order with respect to
interaction and describe current at large $g$ qualitatively.
Diagrams with larger number of Coulomb lines either give small
corrections proportional to powers of $V/\Omega_p$ or correspond
to a renormalization of the vertex which does not change the
results qualitatively.

For the coherent electron tunneling at large voltage drops per
conducting layer ($eV \gg \hbar/\tau$ in the normal state and $V
\gg \Delta$ in superconducting state) this process contributes to
the current only due to the renormalization of the Green's
function that results in the non-zero spectral density spreading
up to large energy. In this limit, assuming $eV \gg \hbar\omega_p$
(but $eV \gg \hbar\Omega_p$) we get the linear I-V curve with
conductivity
\begin{equation}\label{sigma1}
\sigma_1 = \frac{e^2mst_\perp^2}{\pi \hbar^4\Omega_p}
\frac{g}{(1+g_0)^3}\left[ 1+ \frac{S_p}{2\pi p_F^2 }\right],
\end{equation}
where $S_p$ is the area of the two-dimensional Brillouin zone of
metallic layers. Comparing equation (\ref{sigma1}) with the
standard expression for the linear conductivity, $\sigma
=\omega^2_p \tau/4\pi$, one can see that the role of an effective
scattering time in $\sigma_1$ is played by $\frac{\pi}{8\hbar
\varepsilon_F}\left[ 1+ \frac{S_p}{2\pi p_F^2 }\right]$. In
superconductors with d-wave pairing diagram 1a contribute to the
conductivity at $V < \Delta$ due to transitions of quasiparticles
via the superconducting gap \cite{Art-mech}, the role of the
scattering time being played by $\frac{\pi \hbar}{16 \Delta}$
where $\Delta$ is the maximum value of the gap.

The processes given by the diagram 1b describes conductivity due
to plasmon assisted electronic transitions. Its contribution to
the current has a self-explaining structure
\begin{equation}
\begin{array}{l}
\displaystyle
j =
\int dk d{\bf p} d{\bf p}' e^{iks}
\frac
  {4e^3t_\perp^2 \omega_{\bf q}}
  {\pi^3 \hat q_\perp^2 V^2}
\\
\displaystyle
[ n_{\bf p}(1-n_{{\bf p}'})(1+N_{\bf q}) -
  (1-n_{{\bf p}})n_{{\bf p}'} N_{\bf q}
]
\\
\displaystyle
[
  \delta \, (\xi_p - \xi_{p'} + V + \omega_{\bf q}) -
  \delta\,(\xi_p - \xi_{p'}-V + \omega_{\bf q})
]
\end{array}
,
\end{equation}
where $n_{\bf p}$ and $N_{\bf q}$ are Fermi and Planck
distribution functions for electrons and plasmons. Calculating the
integrals we find at $eV \gg \hbar\omega_p, \Delta$ linear
conductivity equal to
\begin{equation}\label{sigma2}
\sigma_2 = \frac{e^2 p t_\perp^2}{3\pi^3 \hbar^5\Omega_p^2}
\frac{g}{(1+g_0)^3}.
\end{equation}
Under our assumptions the first contribution is larger, than the
second one, $\sigma_1 /\sigma_2 \sim \kappa s$.

Note that similar expressions (\ref{sigma1}-\ref{sigma2}) for the
conductivity is valid in the linear regime at frequencies $eV \gg
\hbar\omega_p, \Delta$, since the linear response at frequency
$\omega$ is described by similar diagrams with voltage $V$
replaced by frequency $\omega$.

Now we discuss a relation of our results to the experimental data.
The most published data are related to high-T$_c$ superconductors.
The latter possess specific features which are not taken into
account in our simple model. For example, we do not take into
account possibility of Van Hove singularities in the vicinity of
the Fermi surface, and angle dependence of $t_\perp$. Therefore,
it is difficult to compare our results with the experimental data
directly. However, qualitative consistency with the data is
present. Our model gives qualitative explanation of the
peak-dip-hump structure in the spectral density on the microscopic
basis and provides the mechanism for the high voltage or frequency
conductivity in the stack direction.

The value of the conductivity
of Bi$_2$Sr$_2$CaCu$_2$O$_x$ measured \cite{LB,Kleiner,Yurgens} at
$V > \Delta$ is few times larger than that at small voltages, $V <
\Delta$. The calculated high voltage conductivity contains an
effective scattering rate of the order of the Fermi energy, while
conductivity at $V < \Delta$ the role of the scattering rate is
played by the superconducting gap $\Delta$. For the isotropic
$t_\perp$ this would imply that the conductivity at high voltages
is smaller than at low voltages. But if the interlayer transfer
integral is dominated by the transfer via apex oxygen, as was
assumed in Ref.~\onlinecite{t}, then the relation between these
conductivities may become the opposite, because conductivity at
small voltages is dominated by the quasiparticles near the nodes
of the superconducting gap where $t_\perp$ is minimal \cite{t},
while to the conductivity at $V > \Delta$ contribute
quasiparticles with any momentum directions, the conductivity
being dominated by directions with large values of the transfer
integral. Thus though we cannot give quantitative description of
the conductivity within our simple model, the results do not
contradict to the experimental data.

We are grateful to K.E. Nagaev and A.G. Kobel'kov for useful
discussions. This work was supported by project 01-02-17527 of
Russian Foundation for Basic Research, and by project 96053 of
Russian program on superconductivity.


\begin{thebibliography}{20}
\bibitem{LB} Yu.\,I. Latyshev, T. Yamashita, L.\,N. Bulaevskii et al., Phys. Rev. Lett. {\bf 82}, 5345 (1999).
\bibitem{Jap} M. Suzuki, T. Watanabe, and A. Matsuda , Phys. Rev. Lett. {\bf 82}, 5361 (1999).
\bibitem{Yurgens} V.\,M. Krasnov, A. Yurgens, D. Winkler et al., Phys. Rev. Lett. {\bf 84}, 5860 (2000).
\bibitem{Kleiner} R. Kleiner and P. M\"{u}ller, Phys. Rev. B {\bf 57}, 14518 (1998).
\bibitem{freq} S. Tajima, J. Sch\"{u}tzmann, S. Miyamoto et al. Phys.  Rev.  B, {\bf 55}, 6051 (1997).
\bibitem{Ta} B. Ruzicka, L. Degiorgi, H. Berger et al., Phys. Rev. Lett. {\bf 86}, 4136 (2001).
\bibitem{ARPES} D.\,S. Dessau, B.\,O. Wells, Z.-X. Shen {\em et al.}, Phys. Rev. Lett. {\bf 71}, 2781 (1993).
\bibitem{Huang} Qiang Huang, J.\,E. Zasadinski, K.\,E. Gray et al., Phys. Rev. B, {\bf 40}, 9366 (1989).
\bibitem{IM} L.\,B. Ioffe and A.\,J. Millis, Phys. Rev. B, {\bf 61}, 9077 (2000).
\bibitem{Norman} M.\,R. Norman, H. Ding, Phys. Rev. B, {\bf 57}, R11089 (1998); M. Eschrig, M.\,R. Norman, Phys. Rev. Lett. {\bf 85}, 3261 (2000).
\bibitem{Chub} A.V. Chubukov and D.K. Morr, Phys. Rev. Lett. {\bf 81}, 4716 (1998)
\bibitem{plth} T. Mishonov, Phys.  Rev.  B, {\bf 44}, 12033 (1991); {\bf 50}, 4004 (1994); S.\,N. Artemenko and A.\,G. Kobel'kov, Pisma Zh. Eksp. Teor. Fiz. {\bf 58}, 435 (1993) [JETP Lett. {\bf 58}, 445 (1993)]; M. Tachiki, S. Koyama, and M. Takahashi, Phys. Rev.  B, {\bf 50}, 7065 (1994).
\bibitem{plex} Ophelia K.\,C. Tsui, N.\,P. Ong, Y. Matsuda et al., Phys. Rev. Lett. {\bf 73}, 724 (1994); Y. Matsuda, M.\,B. Gaifullin K. Kumagai et al., Phys. Rev. Lett. {\bf 75}, 4512 (1995).
\bibitem{Art-mech} S.\,N. Artemenko, Pis'ma Zh. Eksp. Teor. Fiz. {\bf 70}, 526 (1999) [JETP Lett. {\bf 70}, 516 (1999)].
\bibitem{t} O.\,K. Andersen, A.\,I. Liechtenstein, O. Jepsen et al., J. Phys. Chem. Solids {\bf 56}, 1573 (1996).

\end{thebibliography}
\end{document}